# IMPROVING THE JET RECONSTRUCTION WITH THE PARTICLE FLOW METHOD; AN INTRODUCTION

JEAN-CLAUDE BRIENT

*Laboratoire Leprince-Ringuet, Ecole polytechnique,*
*Route de Saclay, 91128 Palaiseau, France*

At the future e+e− linear collider, the reachable physics will be strongly dependent on the detector capability to reconstruct high energy jets in multi-jet environment. At LEP, SLD experiments, a technique combining charged tracks and calorimetric information has been used to improve the jet energy/direction reconstruction. Starting from this experience, it has been proposed to go from partial individual particle reconstruction to complete (or full) individual reconstruction. Different studies have shown that the reachable resolution is far beyond any realistic hope from calorimetric-only measurement.

## 1. Introduction

At the future e+e− linear collider, the production of multi boson events will be of major importance for the physics output of the machine and consequently the separation between boson species, Z, W and Higgs, will be crucial. The LEP experiments have shown that a good signal/noise ratio for di-boson WW, ZZ decaying to 4 jets final state could be reached with an excellent purity. It must also be the goal for the new machine, where a maximal use of the luminosity must include the use of bosons decaying to jets. How good the boson separation for the new machine must be, has been a subject of studies for a long time [1]. A known example is given by the separation between Z pair and W pair in the reaction $e^+e^- \rightarrow W^+W^-\nu\nu$ and $e^+e^- \rightarrow ZZ\nu\nu$, where no kinematics fit can be performed and therefore where detector capability is the key parameter. To quantify the impact of the jet energy resolution, a sample of events is first generated using PYTHIA [2]. Visible[*] MC particles are used to reconstruct jets, using the E-improved Jade algorithm, and forced to split the events in 4 jets. Then, writing the jet energy resolution as $\Delta E = \alpha\sqrt{E}$, the energy and direction of the jets are smeared following the value of $\alpha$. The 4 jets are then paired into 2 di-jets following the di-jets mass difference with boson masses. Testing, with this method, the impact of different $\alpha$ values, leads to automatically taking into account the fluctuation of the jet fragmentation, the jet finding and the imperfection of the jet algorithm (particle mismatch). The separation WW, ZZ with the value of α is illustrated on Figure 1. The impact is interpreted as coming from jet pairing and di-jet mass resolution. Quantitatively, for the measurement of the longitudinal coupling of the W boson, the loss of efficiency and purity when going from α=0.3 to 0.6, which is a typical LEP value, is equivalent to a 40% loss of luminosity for 3 years of data taking at 500 GeV.

---

[*] Visible particles are defined as particles with energy above some reasonable threshold (i.e. Pt min for charged tracks at 200 MeV/c), inside the detector fiducial volume and not escaping like neutrino(s).

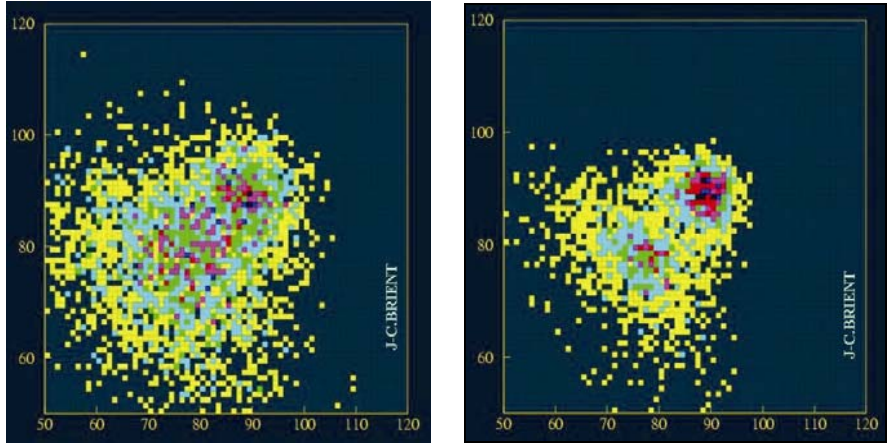

Figure 1: For $W^+W^-\nu\nu$ and $ZZ\nu\nu$ events in 4 jets, distribution in the plane of the two di-jets masses, for α=0.6 (left) and 0.3 (right). The WW and ZZ separation is clearly visible only for α=0.3.

Another study performed on the crucial measurement of the Higgs to WW* branching ratio, leads to a similar effect, with an equivalent loss of 44% of the luminosity for again 3 years of data taking on the precision of the measurement.

## 2. The particle flow paradigm

When looking at the average energy content of a jet with about 65% coming from charged track(s), 26% from photon(s) and about 9% from neutron(s) and neutral hadron(s), it seems natural to use tracker device(s) for the charged tracks energy estimation since the "tracker" device(s) have usually a very good momentum resolution. The calorimeter is therefore dedicated to measure only neutral particle(s). Consequently, it must be able to disentangle the contribution from neutral(s) from the one of charged particle(s). With this method, each particle is reconstructed *individually* like in a bubble chamber, which is the best reconstruction one can imagine. This method is called Particle Flow or PFLOW. A simple calculation using a standard tracker momentum resolution for the charged tracks, a typical ECAL energy resolution of 10% stochastic term for the photon(s) and a typical HCAL energy resolution of 40% stochastic term for neutron(s) and neutral hadron(s), gives a very good jet energy resolution with α=0.14. This value seems far away from any purely calorimetric measurement, even with very good e/h and very good performance on single particles, especially in a 4T magnetic field as needed to contain the machine background. Up to now, only two LEP experiments and SLD take this way to reconstruct the jet energy, at least for the photons part of the jet [3]. It is due to the good segmentation of the ECAL in these experiments. Forgetting the correlation, we can write the jet energy resolution as follow:

$$\sigma^2_{jet} = \sigma^2_{h\pm} + \sigma^2_{\gamma} + \sigma^2_{ho} + \sigma^2_{confusion} + \sigma^2_{threshold} + \sigma^2_{losses} \qquad (1)$$

In this formula, $\sigma_{h\pm}$ is the resolution coming from the charged tracks, $\sigma_{\gamma}$ the one from the photon(s) and $\sigma_{ho}$ the one from neutron(s) and neutral hadron(s). For perfect PFLOW, the contributions stopped there. For a real collider detector, there is also the contribution coming from the mixing of the deposited energy between neutral and debris of the charged hadrons interaction in the calorimeter ($\sigma_{confusion}$), the losses of particles due to imperfect reconstruction ($\sigma_{losses}$) and the threshold of energy for each species which integrate the fluctuation at low energy of the jet fragmentation ($\sigma_{threshold}$). To reach α=0.14, as given above, it is needed to forget the last three terms, and therefore this value can be considered as the ultimate value for the PFLOW method.

Here comes the main difference with the "standard" energy flow method (EFLOW), where the energy is statistically estimated by weighting due to the geometrical overlap of the neutral and charged showers. With charged track(s) and neutral particle(s) close enough to see only a single shower, the neutral particle(s) can be missed and in any case the energy resolution will depend on the knowledge of the energy deposited by the charged tracks in the calorimeter. This statistical approach of the EFLOW has an energy resolution which of course cannot be as good as the PFLOW approach. Taking the example of the photons, in the case of the PFLOW, $\sigma_{\gamma}$ is directly related to the ECAL resolution on e.m. shower. For the EFLOW method, an important contribution to $\sigma_{\gamma}$ comes from the imperfect knowledge of the hadronic interaction of the charged track in the ECAL, with problems like the knowledge of the hadronic energy scale, the hadronic energy resolution, transversal distribution of the hadronic shower, etc…

### 3. How to optimize the PFLOW ?

The best value of α obtained up to now is about 0.6 at LEP/SLD, far away from the perfect reconstruction with α=0.14. It means that the contributions coming the last three terms of the equation (1) are dominant. Studying the effect of the threshold with simulation, the observed effect is modest and a good signal/noise at low energy could further reduce it. The main contributions are therefore coming from the losses and from the confusion. Optimizing the detector for the PFLOW performance leads therefore to minimize the confusion between showers and the losses of particles in the detector. For the confusion, it means to associate as correctly as possible the deposited energy with the particle source and concerning the losses, it means to have the best possible shower–charged tracks separation. This is illustrated by the variable called separability which in case of photons can be expressed like :

$$S_{h\pm/\gamma} \sim BL^2/(R_M \oplus \lambda_{had} \oplus D_P) \qquad (2)$$

Where B is the magnetic field, L is the radius of the ECAL entrance, $R_M$ is the effective Molière radius and $\lambda_{had}$ is the interaction length of the ECAL and Dp is the pad size of the readout in ECAL. Larger is $S_{h\pm/\gamma}$ better is the separation between charged tracks and photons. In this condition, the optimisation goes to a large B-field or even better a large internal radius of the ECAL while a small Molière radius, a small pad size and a large interaction length are obvious. To have a picture of the region of interest for the Molière radius, Figure 2 shows for different physics processes at √s = 800 GeV, the average fraction of the photon energy per event which are closer to some distance of any charged track, i.e. about 10% of the photons energy are within 2cm from the extrapolation of a charged track. It concerns the heart of the jet, and therefore the closest charged hadron to the photons is usually with a large momentum (more than 10 GeV/c). An effective Molière radius smaller than about 2cm is obviously mandatory for this physics processes.

The second point in order to avoid confusion consists in having a three dimensional view of the showers, allowing a good pattern recognition and therefore a good shower separation. It is illustrated on Figure 3, which shows a part of a WW to jets event simulated, using GEANT4, at √s = 800 GeV, for a dedicated calorimeter (i.e. with large segmentation in depth and small pad size). Of course, it means also the development of software working in 3D, for the shower pattern recognition. Dedicated algorithms are under development in many groups involved in the LC detector R&D.

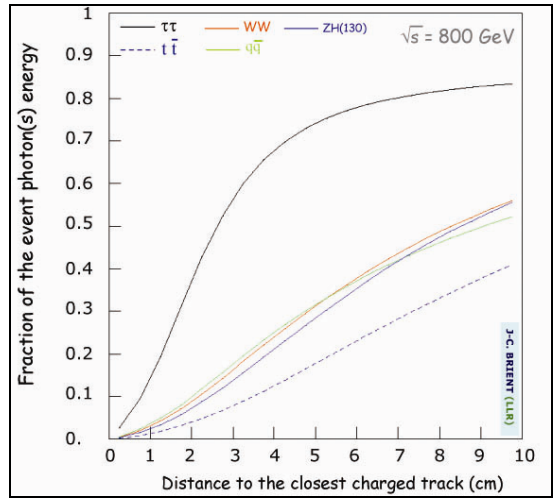

Figure 2: the average fraction of the photons energy for photon closer to some distance of any charged track. The detector and B field are taken from [1]

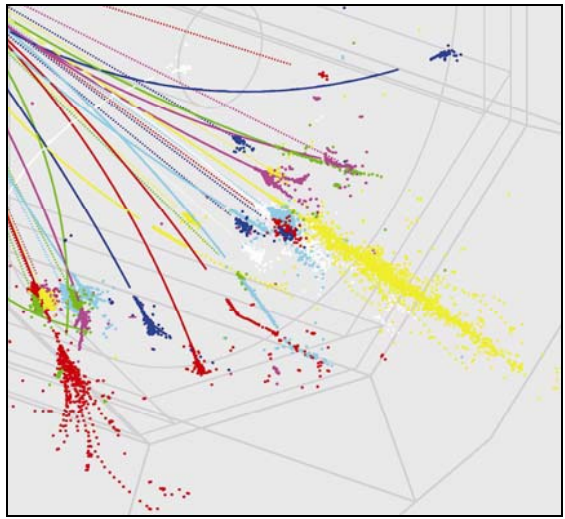

Figure 3: A zoom of the calorimeter with a W decays to jets at √s =800 GeV. The event is simulated using GEANT4 for a large segmentation calorimeter (example taken from CALICE collaboration[4])

Different techniques exist on the way to proceed to do the particle flow in an event. The usual way consists in using information from different devices together, that is tracker, ECAL and HCAL. For example, extrapolation of charged tracks can be used to match or not with clusters in ECAL, and is a basic element for photon(s) reconstruction.

## 4. First results with simulation

Using GEANT4, reactions like $e^+e^- \to Z$ at $\sqrt{s}$=91 GeV or $e^+e^- \to W^+W^-$ at $\sqrt{s}$=800 GeV, with all bosons decaying to jets, have been simulated. ALEPH photon reconstruction package [5] has been adapted to the geometry and segmentation of the proposed CALICE ECAL and HCAL. The result on the photon(s) is impressive. Figure 4 shows, for example, the generated and reconstructed distributions of the photon(s) energy and multiplicity for the W pairs. The excellent agreement shows that the extension of the LEP/SLD technique is possible at higher energy with a suitable detector. Similar results exist also for the neutral hadron(s) and for the full jet reconstruction [7] and it is illustrated on Figure 5, which shows the visible energy distribution at the Z peak for one of the CALICE HCAL options (1cm$^2$ digital readout pad size), and the Z mass distribution for a more classical HCAL option based on scintillator tiles. In both cases, these preliminary studies show that the region of $\alpha$=0.3 could be reached at the Z peak. More works on algorithm are under way to keep this value of $\alpha$ for higher energy domain.

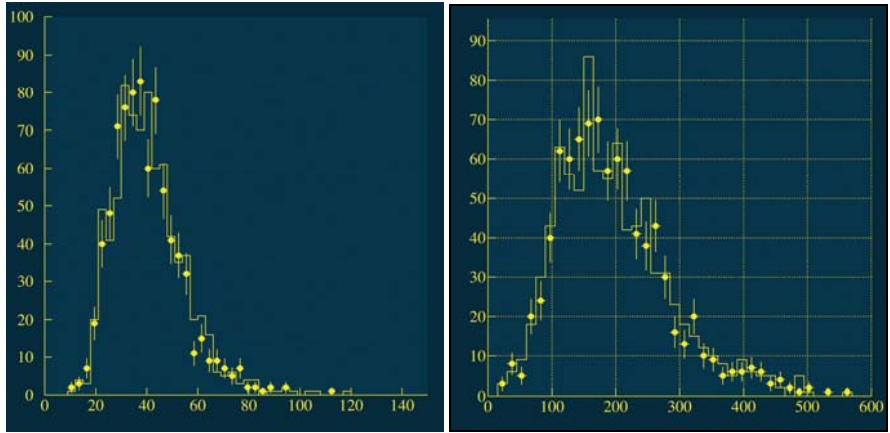

Figure 4: WW decaying to jets at $\sqrt{s}$ =800 GeV. Distribution of the photons energy per event (left) and the number of photons per event (right). Histogram is the generated distribution and dots are the reconstructed one. (Example taken from CALICE collaboration [6])

## 5. Conclusions

The expected physics programme at the TeV $e^+e^-$ linear collider is dominated by multi-bosons processes and therefore by final states with a large multiplicity of jets. The most promising way to improve the jets reconstruction is the extension of the particle flow method initially developed at LEP and SLD. In this case, ultra segmented calorimeters are the best choice for this method. First results based on GEANT4 simulation and preliminary dedicated algorithm lead to factor of two improvement versus LEP/SLD performance reaching the region of $\Delta E_J = 0.3\sqrt{E_J}$, considered as mandatory for the LC physics programme.

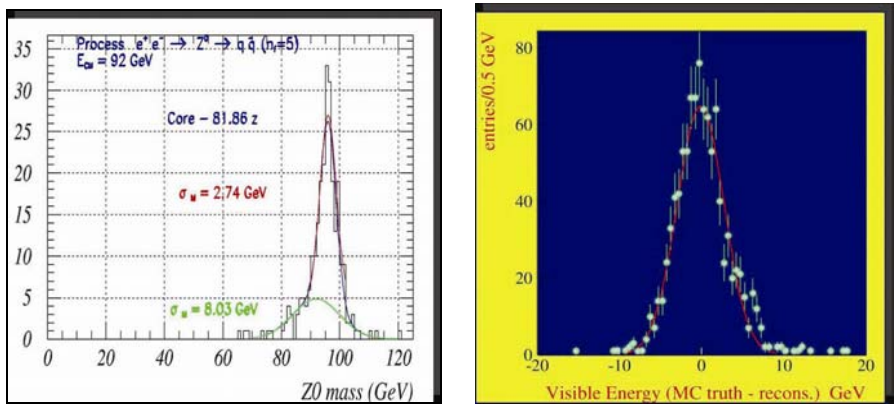

Figure 5: Z decays to jets at $\sqrt{s}$ =91 GeV. On the left, the visible mass distribution with CALICE ECAL and tile HCAL. The core of the distribution corresponds to a resolution of 2.74 GeV. On the right, the visible energy distribution for CALICE ECAL and digital HCAL (readout pad size of 1cm$^2$), with a distribution about Gaussian and standard deviation of 2.9 GeV